\documentstyle[preprint,aps,epsf]{revtex}
\begin{document}
\draft
\preprint{43rd annual MMM conf., Nov. 9-12, 1998}
\title
{Quantum computing and single-qubit measurements using the spin filter effect}
\author{David P.
DiVincenzo\footnote{divince@watson.ibm.com}}
\address{
IBM Research Division, T. J. Watson Research Center,
P. O. Box 218, Yorktown Heights, NY 10598, USA}
\date{\today}
\maketitle
\begin{abstract}
Many things will have to go right for quantum computation to become a
reality in the lab.  For any of the presently-proposed approaches
involving spin states in solids, an essential requirement is that
these spins should be measured at the single-Bohr-magneton level.
Fortunately, quantum computing provides a suggestion for a new
approach to this seemingly almost impossible task: convert the
magnetization into a charge, and measure the charge.  I show how this
might be done by exploiting the spin filter effect provided by
ferromagnetic tunnel barriers, used in conjunction with one-electron
quantum dots.
\end{abstract}
\pacs{1998 PACS: 03.67.Hk, 75.10.Jm, 73.61.-r, 89.80.+h}



\section{Quantum Computing}

In its initial development, quantum computing seemed remote from the
concerns of solid state or magnetic physics.  It emerged initially, in
the early 80s, as an abstract concept for computation that was studied
by theoretical physicists concerned with the foundations of quantum
theory; by the late 80s, it began to be of interest also to a small
group of theoretical computer scientists. As Rolf Landauer has said,
these theorists were much more interested in Hamiltonians and state
vectors than in devices.

But the discoveries of the computer scientists, notably that of Peter
Shor in 1994\cite{Shor}, showed that quantum computing cannot be ignored
as a future technology.  If the bits of a computation can be embodied in
quantum two-level systems (qubits) that would exist and be manipulated
as coherent quantum states, the rules of computation become different;
some calculations which will always be intractable on an ``classical"
computer are possible, and perhaps ultimately easy, on a quantum
one\cite{Divrev}.

It is by now pretty clear what requirements must be fulfilled for a
particular quantum system to realize a quantum computer\cite{5rules},
and we can see now that there is the potential for achieving the
necessary conditions in the solid state.  These rules have been explored
in detail elsewhere\cite{Rolf}, but very briefly they may be summarized
as follows.  A simple quantum system must be available with a small
number of discrete quantum states (ideally two to represent the qubit);
it must be possible to assemble many such qubits and controllably couple
and uncouple them with a repertoire of time-dependent Hamiltonians which
accomplish the action of logic gates; the states of these qubits must
have high quantum coherence, meaning that the state's decoherence time
(during which the state entangles significantly with quantum states in
the environment) must be at least 1000 times longer than the time to
execute a logic gate; it must be possible to set these qubits into their
ground state initially; and it must be possible to measure, reliably and
quickly, the state of each qubit individually.

Much could be said about the quest to satisfy all these requirements.
Efforts are underway to create simple quantum computer gates in many
areas of experimental physics: in atomic spectroscopy, in cavity
quantum electrodynamics, in nuclear magnetic resonance on organic
molecules, and in ion traps.  A description of all of these would take
me outside the scope of the present article, but I would mention one
point on which these diverse areas intersect with the subject of this
conference: in the majority of cases, the qubit is realized by a spin
degree of freedom.  This choice leads to very readily satisfying a few
of the criteria for realizing quantum computation--for one thing, one
obtains automatically a well defined two level system (for spin 1/2
degrees of freedom), and in addition, it is often the case that this
degree of freedom interacts quite weakly with the environment, and
thus has a long decoherence time.

Thus, the thinking that we\cite{LD} and others\cite{Kane} have been
doing about the potential solid-state implementations of quantum
computers have also focussed on spin degrees of freedom.  (This thinking
has not been exclusively confined to spin; I will not touch here on
qubits based on superconducting degrees of freedom\cite{Shnir} or
quasiparticle excitations of an anyonic system\cite{Kitaev}).  Of
course, we have had a long history of thinking about quantum effects in
the dynamics of the magnetization of small magnetic particles and
magnetic molecules.  But this is actually not a good precedent for
quantum computation either, as our focus for many years was on {\em
macroscopic} magnetic effects\cite{ADS}, either MQT (tunneling) or MQC
(coherence).  

While it has been very fascinating that large systems can show any
quantum effects at all, they are not particularly good candidates for
quantum computing, since, being macroscopic, they interact readily
with degrees of freedom in the environment.  Thus, they do not show
the very high degree of quantum coherence that is desirable for qubits
in a quantum computer. I would not say, however, that by going over to
microscopic magnetic degrees of freedom that the problems of achieving
quantum computation become much easier; indeed, the proposals that are
presently on the table involve tremendous extensions beyond the
current experimental state of the art.

But, some consensus seems to be developing about a likely path for
solid state quantum computing with spins.  At least, two serious
proposals that are on the table\cite{LD,Rolf,Kane}, starting from two
very different points of view, have many elements in common. In both,
the spin of a single electron confined to a localized orbital state is
used for quantum computation.  In the Loss-DiVincenzo proposal the
confinement is produced by a quantum dot deep in the Coulomb blockade
regime; in the Kane scheme, the electron is confined in the bound
state of a shallow donor, P in Si.  (The proposals differ in that in
the Kane proposal, the nuclear spin of the P is also employed as a
qubit.)  In both, quantum logic gates are achieved by voltage gates
which manipulate the shape of the electron orbital, changing the
overlap of the electron wavefunctions on neighboring quantum dots (or
P impurities).  Both need effective local magnetic fields to achieve
spin precession of individual selected spins: Loss and DiVincenzo
envision that small local fields are applied directly, Kane imagines
changing the effective (Overhauser) field arising from the hyperfine
nuclear-electronic coupling by distorting the electron orbital.

\section{Single-spin magnetometry}

And, both proposals require that quantum measurements be made at the
single-spin level, and this will be the subject of the rest of this
essay.  It may seem maddeningly glib for theorists to pronounce, ``the
experimentalists shall measure magnetization at the one-Bohr-magneton
level," and expect it to be done.  Believe me, we have little choice;
achieving quantum computation requires no less, much as we wish that
it would.  This theorist, at least, has observed with interest over
the years the Herculean efforts that have been made by
experimentalists to achieve ever more sensitive magnetometers, and
recalls the landmark achievements in SQUID design which made it
possible to push sensitivities down to $10^5\ \mu_B$\cite{ADS}.

How, then, are we to do five orders of magnitude better?  Of course,
the experimentalists have not been inactive on this front, many have
been questing towards this holy grail of single-quantum sensitivity.
Rugar and his collaborators\cite{Rugar} have pushed closer than anyone
else to arrive at this point by mechanical detection means (using a
magnetic force microscope).  But the work has been very hard, and they
have not quite measured a single spin yet.  Optical methods would seem
to offer another approach--it is indeed by optical detection that
single-atom measurements can be done in atomic physics--and the
near-field optical techniques of Moerner\cite{WE} and of Gammon and
coworkers\cite{Gam} would seem to point the way towards another method
of interrogating single quanta.  I will be very interested if this
technique can be pushed to see single spins separated on the
tens-of-nanometers scale.

But in this essay I want to give some details of another completely
different approach which we have proposed\cite{LD,Rolf}, largely
inspired by the mindset of quantum computing, for achieving single-spin
magnetometry by purely electrical means.  The idea which we bring over
from quantum computing to provide a new approach to this problem is that
of coherent transfer of a quantum state from one embodiment to another. 
If it is possible to turn on and off the right kind of Hamiltonian
between two physically different qubits A and B, then it is possible to
swap the states of these two objects: 
\begin{equation}
|\psi\rangle_A\otimes|0\rangle_B\stackrel{swap}{\rightarrow}
|0\rangle_A\otimes|\psi\rangle_B. 
\end{equation} 
In quantum computing an interaction which performs this function is
referred to as a ``swap gate."  It is quite trivial from the point of
view of computing (it just moves bits around, it doesn't actually
perform any `useful' logic operation on them), but it can mean
something quite important for the measurement problem.  Generally
speaking, it may permit one to turn a very hard-to-measure quantum A
into an easier-to-measure quantum B.

It is probably important to add a reminder here that a measurement at
the quantum level has a qualitatively different character than the
usual magnetization measurement, in that it has an unavoidably
``digital'' character\cite{TD}.  In short, if the wavefunction of the
single spin is
$\psi=\alpha|\!\uparrow\rangle+\beta|\!\downarrow\rangle$, then a
successful von Neumann measurement should announce the outcome ``up''
with probability $|\alpha|^2$, and outcome ``down'' with probability
$|\beta|^2$.  The laws of quantum mechanics do {\em not} permit a
measurement to obtain the value of $\alpha$ itself (at least, not
using just one shot of the measurement), as we would expect if we
naively extrapolate the usual magnetization measurement down to the
Bohr-magneton level.  We will see how the measurement proposed below
conforms to this ``digital'' rule.

\cite{LD} and \cite{Kane} both invoke as the preamble to a quantum
measurement such a reimbodiment, in particular, one which maps an
electron {\em spin} quantum into an electron {\em charge} (or orbital)
quantum state.  This is a very desirable transformation, since there are
known methods for performing electron-charge measurements with exquisite
sensitivity, reportedly at sensitivity levels of $10^{-8}$ of one
electron, using either a single-electron transistor\cite{Devoret} or a
quantum point contact\cite{Moty}.  (Kane's scheme\cite{Kane} requires
concatenating this with another swap, one which takes a nuclear spin
state to an electron spin state.  I will not discuss this step here.)

How is this swap, involving a change of the carrier of the quantum bit
state, to be achieved?  We are helped some by the fact that since the
transformation is to be immediately followed by a genuine measurement,
we do not have to be extremely fussy about the maintenance of full
quantum coherence during the swap; in fact, dephasing in the basis of
the measurement will be entirely permissible.  This permits us to
consider incoherent dynamics, such as tunneling, to accomplish the
desired swap.

\begin{figure}
\epsfxsize=15cm
\epsfbox{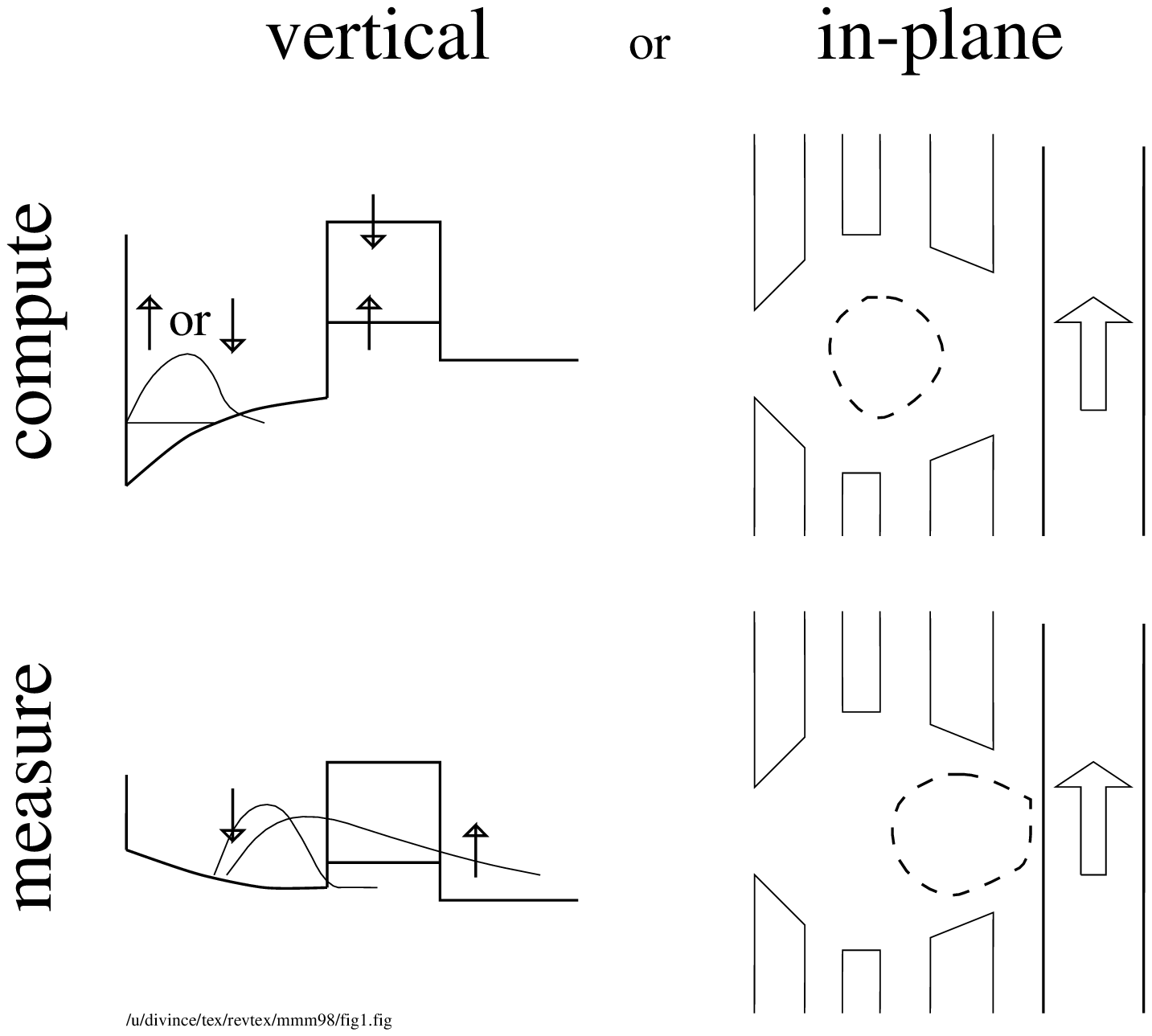}
\caption{Schematics of the spin-filter quantum measurement.  Possible
vertical and in-plane structures are shown.  In the compute phase the
electron wavefunction in the quantum dot is held far away from the
ferromagnetic insulator barrier.  In the in-plane structure, the classical
turning point of the quantum dot potential is shown dashed.  In the measure
phase the gate potentials are changed so that the electron wavefunction
is pressed up against the ferromagnetic barrier.  Because of the exchange
splitting of the conduction electron barrier in the ferromagnet, spin up
will tunnel through the barrier easily while spin down will not.}
\label{fig1}
\end{figure}

Now I finally come to the specific ideas, which we have touched upon
previously\cite{Rolf,LD}, for how to measure the spin of one electron.
Figure \ref{fig1} illustrates two versions of the basic idea.  In
both, a single electron is held inside a quantum dot, with its spin
state being controlled by quantum gate operations; the control devices
needed to perform these are not shown here\cite{LD}.  We imagine these
dots to be confined regions within a heterostructure quantum well
formed in GaAs or Si/Ge structures, although other quantum dot
implementations (for examples, those obtained by self-assembled
growth) are possible too.

The key element of these structures is a layer, lying either above the
quantum dot in the vertical structure, or in a trench interrupting the
two-dimensional quantum well in the in-plane structure, of an insulating,
magnetized material.  While insulating ferromagnets are not so common
(the spins in insulators tend to order antiferromagnetically), some are
well known and have been studied extensively.  For example, it was shown
long ago\cite{SvM} that the europium chalcogenides, in particular EuSe
and EuS, can be grown in thin films which exhibit a strong ``spin filter''
effect\cite{MHGM88,HMM90,MMH93}, in which carriers of one spin orientation
tunnel through the barrier preferentially between two metal electrodes.
In favorable cases\cite{MMH93} the spin polarization in tunneling has
exceeded 99\%.  

We propose to exploit the ``spin filter'' property of such a barrier
to accomplish the swap necessary to do the quantum measurement.  As
the figure shows, in the ``measure'' phase a gate voltage (or several
of them) is changed so that the electron wavefunction is pressed
against the magnetic barrier.  The magnetization in the barrier
produces an exchange splitting of the conduction band (known to be
0.36eV for EuS in zero applied magnetic field\cite{MHGM88,EuO}), so
that the barrier height for the two spin directions differs by this
amount.  If the band lineups are such that the low barrier is
$\ll$0.36eV as indicated in Fig. \ref{fig1}, then conditions can be
achieved such that spin-up has a very high probability of tunneling
through the barrier, while spin-up has virtually none.

Thus, the desired interconversion has taken place.  The measurement of
spin up or down is converted to a measurement of whether the electron
is on the left or the right of the barrier, or alternatively, whether
the electron can pass repeatedly though the barrier or not.  The
latter suggests an a.c. capacitance measurement, as proposed by
Kane\cite{Kane}.  In either case this becomes a problem in
electrometry; many measurements of electric charge at the
single-electron sensitivity level have now been done, either with
single electron transistors\cite{Devoret}, or with quantum point
contacts\cite{Moty}.  We will not do any analysis of the requirements
of this measurement here, but we note that an excellent study of the
parameters required in an single-electron-transistor measurement for
this application has been given in the context of superconducting
quantum-box qubits\cite{SS}.

Many of the parameters of the europium compounds which would be
relevant to this measurement proposal are well known; EuO\cite{EuO}
and EuS are low temperature ferromagnets (Curie temperatures of 69K
and 16.6K), while EuSe becomes ferromagnetic in a moderate applied
field.  The energy gaps for EuO and EuS are 1.1eV and 1.65eV, with
localized $f$ states forming the valence band; the conduction band is
formed from delocalized $d$ states, which makes the description of the
tunneling of band electrons through the barrier of Fig. \ref{fig1}
valid.  Of course, much is {\em not} known that would be needed for
the present proposal: in particular, the compatibility of these
magnetic materials with conventional semiconductors like GaAs or Si or
Ge is apparently unknown.  The band lineups, epitaxy and quality of
the interfaces, the nature of interface states, would all have to be
understood for this idea to work.  It may be that some other magnetic
insulator, perhaps a transition metal oxide, would better fulfill the
requirements of this proposal.  I hope that workers in magnetic
materials will be interested in conducting this possibly arduous
search for a compatible materials system; the payoff for finding such
a system would be undeniably great for physics.

\acknowledgments
Thanks to Stephan von Moln{\'a}r, John Slonczewski, Daniel Loss, and Guido
Burkard.

\end{document}